\documentclass[12pt]{iopart}
\usepackage{cite}
\usepackage{graphicx}
\usepackage{lipsum} 
\usepackage{acronym,xspace}
\usepackage{macros}

\usepackage{color}

\usepackage{lineno}

\usepackage{array, multirow}   
\usepackage{booktabs}
\usepackage{tabularx}
\usepackage{rotating}
\usepackage{lscape} 
\usepackage{geometry}
\usepackage{color, colortbl}
\definecolor{Gray}{gray}{0.9}
\usepackage{makecell}

\usepackage{hyperref}
\hypersetup{
    colorlinks=true,
    linkcolor=blue,
    filecolor=magenta,      
    urlcolor=cyan,
}

\DeclareFontFamily{OT1}{pzc}{}
\DeclareFontShape{OT1}{pzc}{m}{it}{<-> s * [1.10] pzcmi7t}{}
\DeclareMathAlphabet{\mathpzc}{OT1}{pzc}{m}{it}

\usepackage{iopams}  

\begin{document}

\title[PyChChoo]{Application of a new transient-noise analysis tool for an unmodeled gravitational-wave search pipeline}

\author{Kentaro Mogushi}

\address{Institute of Multi-messenger Astrophysics and Cosmology, Missouri University of Science and Technology, Physics Building, 1315 N.\ Pine St., Rolla, MO 65409, USA}

\vspace{10pt}
\begin{indented}
\item[]\today
\end{indented}

\begin{abstract}
Excess transient noise events, or ``glitches'', impact the data quality of ground-based \ac{GW} detectors and impair the detection of signals produced by astrophysical sources. Identification of the causes of these glitches is a crucial starting point for the improvement of \ac{GW} signal detectability. However, glitches are the product of linear and non-linear couplings among the interrelated detector-control systems that include mitigation of ground motion and regulation of optic motion, which generally makes it difficult to find their origin. We present a new software called \tool which uses time series recorded in the instrumental control systems and environmental sensors around times when glitches are present in the detector's output to reveal essential clues about their origin. Applying \tool on the most adversely affecting glitches on \textit{background} triggers generated by one of unmodeled GW detection pipelines called coherent WaveBurst (cWB) operated in the data from the LIGO detectors between January 1st, 2020 and February 3rd, 2020, we find that 80\% of triggers are marked as either being vetoed or unvetoed in common between our analysis and the current LIGO infrastructure. 
\end{abstract}

\section{Introduction}\label{introduction}
The dawn of gravitational-wave (\ac{GW}) astronomy was opened with the first direct detection of a \ac{GW} signal produced from a \ac{BBH} merger \cite{Abbott:2016GW150914} on September 14$^{{\rm th}}$, 2015. 

During the first and second observing runs of LIGO \cite{TheLIGOScientific:2014jea} and Virgo \cite{TheVirgo:2014hva}, nine additional \ac{BBH} mergers and a \ac{BNS} merger were detected with high confidence\cite{LIGOScientific:2018mvr}. Furthermore, \numthird events were observed with high confidence during the first half of the third observation run\cite{Abbott:2020niy}. The detection rate was approximately 1 per week. 

In order to detect \ac{GW} signals, the ground-based \ac{GW} detectors must be extremely sensitive, causing them to become susceptible to instrumental and environmental artifacts \cite{LIGOScientific:2018mvr}. In particular, transient noise artifacts, or {\it glitches} may mimic \ac{GW} signals in their morphology so that it is crucial to differentiate if trigger events identified by \ac{GW} detection pipelines are astrophysical or terrestrial in origin to reduce false detections.

The initial and essential task to identify an event trigger as a glitch is to understand the origin of the glitch. Glitches are, however, the product of linear and non-linear coupling among the interrelated detector-control systems that include mitigation of ground motion and regulation of optical motion, which typically makes it difficult to find their origin. Clues of the origin may be recorded in some of around fifty thousand auxiliary \textit{channels} such as instrumental sensors and environmental monitors. Because the number of channels is numerous, the task to find the clues is typically made by automated software packages.    

LIGO-Virgo collaboration has been using software engines that find the statistical correlation between the excess power recorded in the auxiliary channels and glitches present in the detector's output. An algorithm called \ac{UPV} \cite{Isogai:2010zz(UPV)} finds the statistical correlation using the percentage of the number of the excess power events identified in each of the auxiliary channels in coincidence with glitches in the detector's output, relative to the total number of excess power events. As a consequence, \ac{UPV} vetoes time periods that have a high correlation factor. Similarly, \ac{hVETO} \cite{Smith:2011an(hVETO)} uses a coincidence statistic to find the correlation while minimizing the vetoed time as much as possible. For finding the correlation for a single glitch, Pointy Poisson \cite{Essick:2020cyv} uses a statistical confidence level that can reject the chance-coincidence hypothesis estimated from the excess power events in the longer time window. iDQ \cite{Biswas:2013wfa} calculates the probability that glitches are present in the detector's output as a function of time, inferred from excess power in the auxiliary channels. 

In this paper, we present a new software (publicly accessible in \href{https://git.ligo.org/kentaro.mogushi/origli}{GitLab}) called \tool  (``Python-based glitCh Characterization tool'') designed to identify the clue of the origin of glitches in \ac{GW} detectors and remove the effect in the detector's output. Using a set of glitches, \tool conditions the time series recorded in the auxiliary channels around the glitch-time and then counts the fraction of frequency bins above a threshold in a given frequency band in order to quantify the excess power in coincidence with the glitch. To identify highly correlated channels (so-called \textit{witness} channels), \tool uses the probability showing the loudness of the excess-power measure in the glitch set compared with the measure in another set which is created with randomly selected timestamps when the detector's output is quiet. After witness channels are identified, removable glitches are determined by the probability that the excess-power measure for a glitch belongs to the glitch set.          

The most novel feature of this algorithm is that it can be used as a ``targeted'' approach. To understand the origin of a particular population of glitches having their specific characteristics denoted such as the peak frequency, \ac{SNR} and/or time-frequency morphology, a user can choose the list of those glitches for running \tool. Instead of providing a single channel that is the most significantly correlated, the algorithm can find multiple channels to help a thorough understanding of potential unknown physical couplings inside instruments. Besides, the ultimate goal of \ac{GW} searches is to detect more signals. Not all glitches are adversely affecting \ac{GW} detection pipelines. Therefore, studying all glitches present in the detector's output that typically is made with \ac{UPV} and \ac{hVETO} might introduce redundant removal time periods. To compensate for this issue, only adversely affecting glitches for a \ac{GW} pipeline can be chosen for running this algorithm. We demonstrate the ``targeted'' approach using triggers with high-ranking statistics that are generated by one of unmodeled \ac{GW} detection pipelines. Finding the witness channels is particularly beneficial for unmodeled \ac{GW} detection pipelines \cite{Klimenko:2008fu,Klimenko:2015ypf, Sutton:2009gi} because they are more susceptible to glitches than matched-filter pipelines \cite{Nitz:2017svb, Sachdev:2019vvd} by their design. Conversely, they have potential capabilities to detect \ac{GW} signals with unknown waveforms or signals empowered by unknown sources.

\section{Software architecture} \label{architecture}

\tool aims to identify the essential clues of the origin of glitches in the detector's output and remove the effect of those glitches. A set of glitches can be selected from any \acp{ETG} or glitch databases, e.g., Omicron \cite{ROBINET2020100620}, pyCBC-live \cite{Nitz:2018rgo}, the database created based on Gravity Spy \cite{Zevin:2016qwy}, or user-defined glitches on demand. Those glitches can be further down-selected based on their characteristics such as peak frequencies, \acp{SNR}, and/or particular glitch-class, etc. 

\subsection{Quantify excess power}\label{quantify_excess_power}
In order to identify the origin of glitches driven from terrestrial disturbances, a set of system control sensors and environmental monitors that do not causally follow from the detector's output (so-called \textit{safe} auxiliary channels) is used. To identify safe channels, LIGO uses \ac{hVETO} \cite{Smith:2011an(hVETO)} and Pointy Poison \cite{Essick:2020cyv}.

Witness channels are expected to show excess power in coincidence with a glitch detected in the detector's output. Those channels might record excess power in different frequency bands. Also, a measure quantifying the excess power depends on a provided time window used to calculate itself. To account for these dependencies on excess-power values, \tool uses the one-sided \ac{ASD} that indicates the noise amplitude. The one-sided \ac{ASD} is a square root of the one-sided \ac{PSD} $S(f)$ defined as
\begin{equation}\label{eq:psd}
    \frac{1}{2}\delta(f-f')S(f)=\left<\tilde{n}(f)\tilde{n}(f')\right>\,,
\end{equation}
where the brackets $\left<\cdots\right>$ denote an ensemble average over noise realizations \cite{Cutler:1994ys}, and $\tilde{n}(f)$ and $\tilde{n}(f')$ are the Fourier transforms of the time series $n(t)$ at the frequencies $f$ and $f'$, respectively. Using the \ac{ASD} of the time series recorded in a set of safe channels in the \textit{quiet} time, we define the \textit{stationarity} upper threshold\acused{SUT} (\ac{SUT}) as follows. 

 The \ac{SUT} is obtained from the time series when no glitches are present in the detector's output. We consider Omicron triggers \cite{ROBINET2020100620} with \ac{SNR} $\leq 5.5$ to be the absence of glitches. To calculate values of \ac{SUT}, \tool first selects random timestamps during the quiet period and then chooses time windows for each of the timestamps by randomly selecting durations ($t_{d,s}$) which are log-uniformly distributed between the minimum duration ($t_{d,min}$) and the maximum duration ($t_{d,max}$). The value of $t_{d,min}$ is chosen to be 0.02 seconds because of the computational requirement for the \ac{ASD} calculation in \textsc{GWpy} \cite{2020SciPy-NMeth}. The value of $t_{d,max}$ is typically chosen to be 35 seconds where the majority of glitches (82\% of glitches in the Gravity Spy glitch-database in the \ac{O2}) have durations less than this value. After setting the time window for each of the timestamps, the \ac{ASD} is calculated using time series in this time window. This \ac{ASD} is subsequently normalized with the median value of \acp{ASD} of overlapping periodograms with a single \ac{FFT} duration of $t_{d,s}$ in the time window of 128 seconds spanning around the timestamp. 

\begin{figure}[ht!]
    \begin{minipage}{0.5\hsize}
  \begin{center}
   \includegraphics[width=1\linewidth]{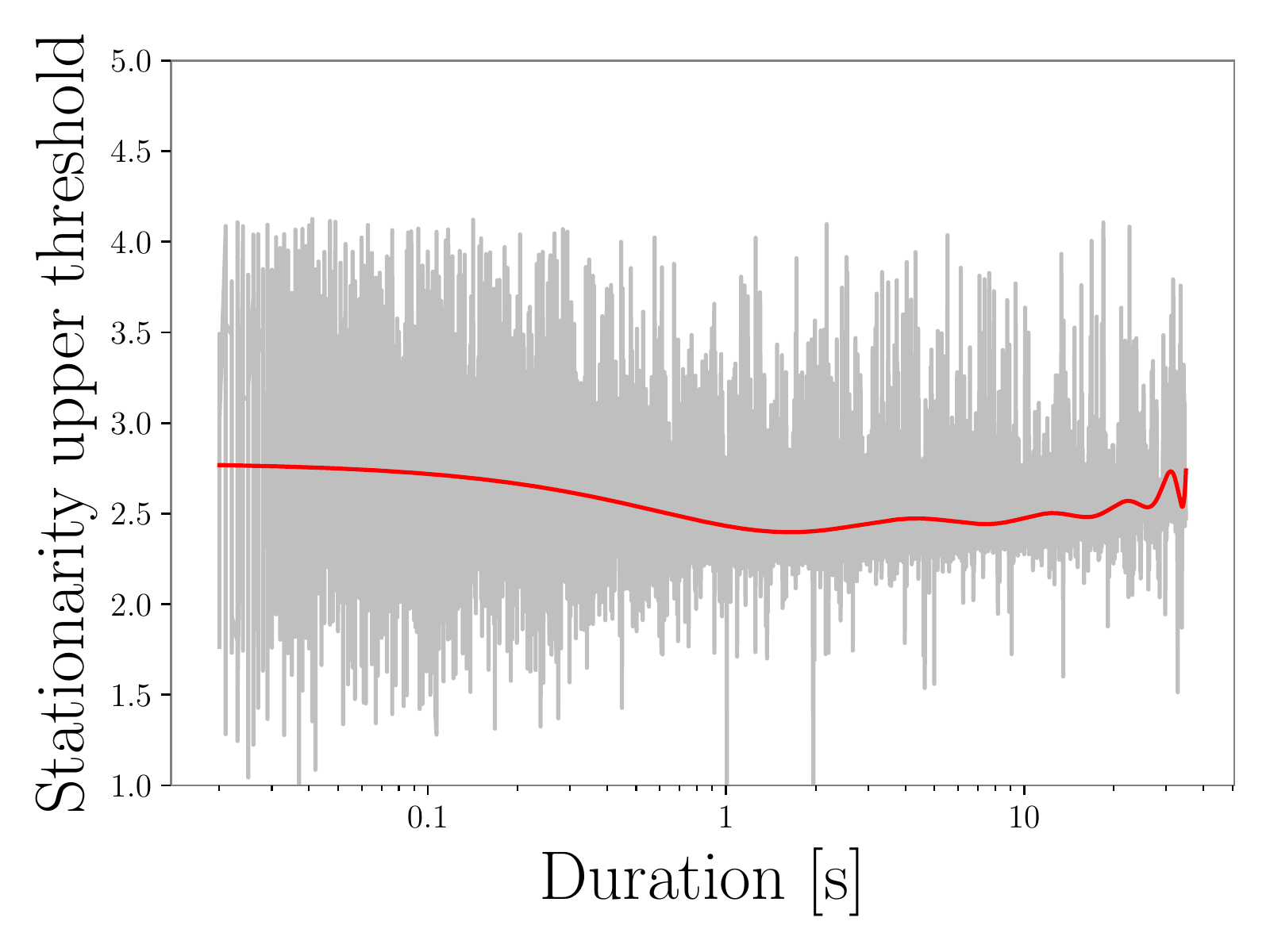}
  \end{center}
    \end{minipage}
    \begin{minipage}{0.5\hsize}
  \begin{center}
   \includegraphics[width=1\linewidth]{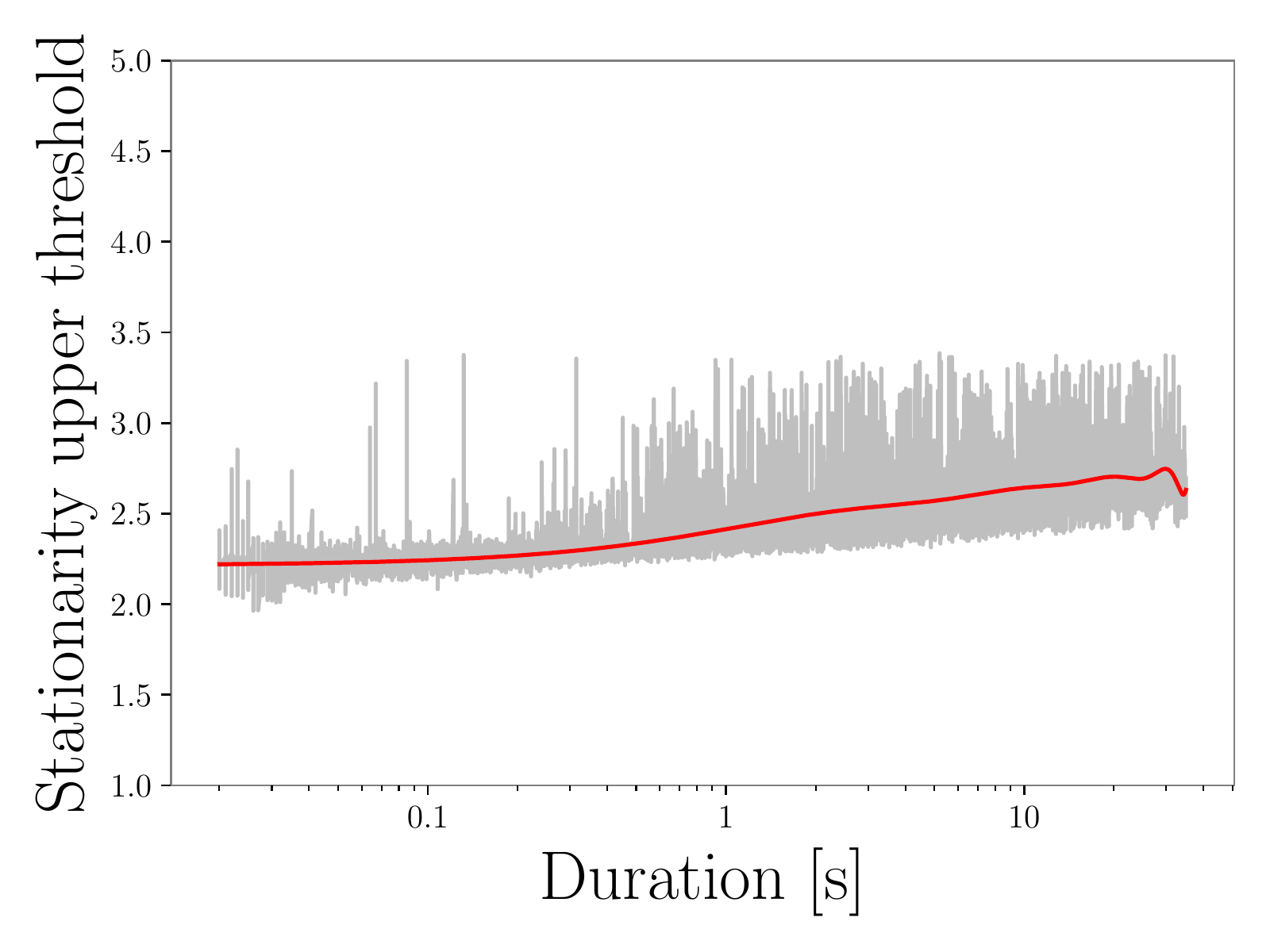}
  \end{center}
    \end{minipage}
\caption{Stationarity upper threshold (\ac{SUT}) of two representative channels: L1:ASC-SRC2\_Y\_OUT\_DQ in 1-128 Hz (left) and L1:PEM\-EX\_MAG\_VEA\_FLOOR\_Y\_DQ in 2048-4096 Hz (right) as a function of duration. The red curves are the polynomial best fits with the degree of 10 obtained from the gray curves that denote the values \ac{SUT} for 8010 different random samples. Outlying values of \ac{SUT} outside of the median absolute error with a 6-$\sigma$ are removed.}
\label{fig:sut}
\end{figure}

Using a set of timestamps, the value of \ac{SUT} is defined as a 3-$\sigma$ standard deviation above the mean value of the normalized \ac{ASD} for each channel in a given frequency band with a given duration. To obtain values of \ac{SUT} for any durations, \tool interpolates \ac{SUT} as a function of duration for each channel in a given frequency band. We find that the polynomial best fit with the degree of 10 while removing the outliers outside of the median absolute error with a 6-$\sigma$ is suitable. Figure \ref{fig:sut} shows the interpolated \ac{SUT} as a function of duration for the two representative channels in particular frequency bands. The interpolated \ac{SUT} are saved and to be used to evaluate glitches like the following.      

For each glitch, \tool conditions the time series recorded in each of the safe channels in the time window:
\begin{equation}
\label{eq:on-sourcewin}
W = \left[ t_g - \alpha t_{d}\,, t_g + (1-\alpha)t_{d}\right]\,,
\end{equation}
where $t_g$ is the time of a glitch, $t_d$ is its duration, and $\alpha$ is a fraction of the duration before $t_g$. A value of $\alpha=0.5$ sets the window to be evenly spanned around the glitch time $t_g$. A value of $t_d$ can be chosen to be the duration provided by an \ac{ETG} or manually selected if needed. Witness channels are expected to record excess power in particular frequency bands in this window when the glitch is present in the detector's output. 

To quantify the excess power for a glitch, \tool firstly calculates the \ac{ASD} of the time series in the window $W$, secondly normalize the \ac{ASD} with the median value of \acp{ASD} which are obtained from the time series in the 128-second window spanning evenly around $t_g$, finally counts the fraction ($q$) of the frequency bins above the value of \ac{SUT} in a given frequency band. Because the sampling frequencies differ between channels (from 256 Hz to 16384 Hz), the lower and/or upper bounds of a chosen frequency band can be greater than the Nyquist frequency of some channels. When only the upper bound is above the Nyquist frequency, we use the frequency bins up to the Nyquist frequency for calculating $q$. When the lower bound is above the Nyquist frequency, we define the value of $q$ to be zero.

\subsection{Probabilistic insight} \label{probabilictic_insight}
After a set of glitches (hereafter called target set) is quantified with values of $q$ from each of the safe auxiliary channels in different frequency bands, the probabilistic measure is used to identify witness channels. Channels with large values of $q$ in the target set could have large values of $q$ during the absence of glitches as well. Channels which record excess power regardless of the presence of glitches indicate no or mild correlation with the glitches. The probabilistic measure accounts for this factor to identify witness channels.

To identify witness channels, \tool compares the target set with another data set (null set). The null set is created by randomly generating time periods with durations being distributed as that of the target set, and then selecting only the subset of these time periods that do not overlap with any glitches being present in the detector's output. We typically consider Omicron triggers with \ac{SNR} $\leq 5.5$ to be the absence of glitches. Because the null set represents the data set when the detector's output is quiet, channels with large values of $q$ in the null set imply no or mild correlation with the targeted glitches.

Witness channels are expected to show a larger number of samples with greater values of $q$ in the target set than the null set. To formulate this manifestation, we consider the distributions $\mathpzc{t}(q)$ and $\mathpzc{n}(q)$ of $q$ in the target and null sets, respectively. The probability that the target set has values in the small interval $(q, q +\Delta q)$ is $\mathpzc{t}(q)\Delta q$, while the probability that the null set has values less than $q$ is $\mathpzc{N}(q) = \int^{q}_0 \mathpzc{n}(q') dq'$, which is known as the cumulative distribution. Multiplying $\mathpzc{t}(q)\Delta q$ and $\mathpzc{N}(q)$ gives the probability of the pair of the above two situations occur in unison. Summing $\mathpzc{t}(q)\Delta q \mathpzc{N}(q)$ over the range of all possible values of $q$, formulated as
\begin{equation}
\label{eq:p_greater}
    p_g = \int_0^1 \mathpzc{t}(q)\mathpzc{N}(q) dq\,,
\end{equation}
is the probability that arbitrary values of $q$ in the target set are greater than values in the null set. Channels with $p_g\lesssim0.5$ imply chance coincidences. Channels with $p_g \sim 1$ indicate evidence of being the witness for the glitches. Therefore, \tool uses $p_g$ to identify the witness channels. In experiments, only a finite number of samples can be obtained. To compensate for this experimental limit, a continuous distribution is preferred to make a robust measure for Eq. (\ref{eq:p_greater}). Because values of $q$ are bounded between 0 and 1, a candidate distribution for $q$ is a Beta distribution. The shape of the Beta distribution is obtained with the first and second moments estimated from the measured samples. 

After witness channels are identified, the effect of the glitches on the \ac{GW} detection pipelines can be mitigated. The simple and standard procedure is to veto the time periods of glitches that are correlated with witness channels. For a glitch, a value of $q$ obtained from the witness channel implies either of two mutually exclusive situations: a value of $q$ follows the target set with being greater than values in the null set; or a value of $q$ follows the null set with being greater than values in the target set. Thus, the probability that $q$ belongs to the target set is given as  
\begin{equation}
\label{eq:p_belong}
    p_v = \frac{\mathpzc{t}(q)\mathpzc{N}(q)}{\mathpzc{t}(q)\mathpzc{N}(q) + \mathpzc{T}(q)\mathpzc{n}(q)}\,,
\end{equation}
where $\mathpzc{T}(q)$ is the cumulative distribution of $q$ in the target set. A value of $p_v \sim 1$ indicates evidence of a strong correlation between excess power in the witness channel and the glitch. \tool uses $p_v$ as a veto criterion.

\section{Software validation} \label{validation}
For validating \tool's performance to identify witness channels, we use a class of glitches with the known instrumental origin that was identified during \ac{O2}. The \ac{L1} detector was contaminated with a new class of glitches between February 9th, 2017 and April 10th, 2019 due to the magnetic coupling between the magnetic field produced from electronics racks and the detector's internal components such as cables, connectors, and actuators \cite{Cavaglia:2018xjq}. These glitches were short-lived spikes with a duration of $\sim 0.3$ seconds and appeared in the frequency band of $\sim$ 50-60 Hz in the detector's output. \ac{hVETO} \cite{Smith:2011an(hVETO)} identified a series of coincident excess power in auxiliary channels in the Physical Environmental Monitor (PEM) mains voltage monitor (MAINSMON) of the Electronics Bay (EBay) in the X-arm end station (EX) as well as in the EX magnetometers. The follow-up study conducted by Ref. \cite{Cavaglia:2018xjq} using the machine-learning-based tools called \textsc{Karoo GP} \cite{DBLP:journals/corr/abs-1708-03157} and \textsc{Random Forest} \cite{Breiman:2001hzm}, identified a physically coupled channel called ISI-ETMX\_ST1\_BLND\_Z\_T240\_CUR\_ IN1\_DQ in the active seismic isolation internal to the vacuum system (ISI) in addition to the EX magnetometer channels. 

\begin{figure}[ht!]
    \begin{minipage}{0.5\hsize}
  \begin{center}
   \includegraphics[width=1\linewidth]{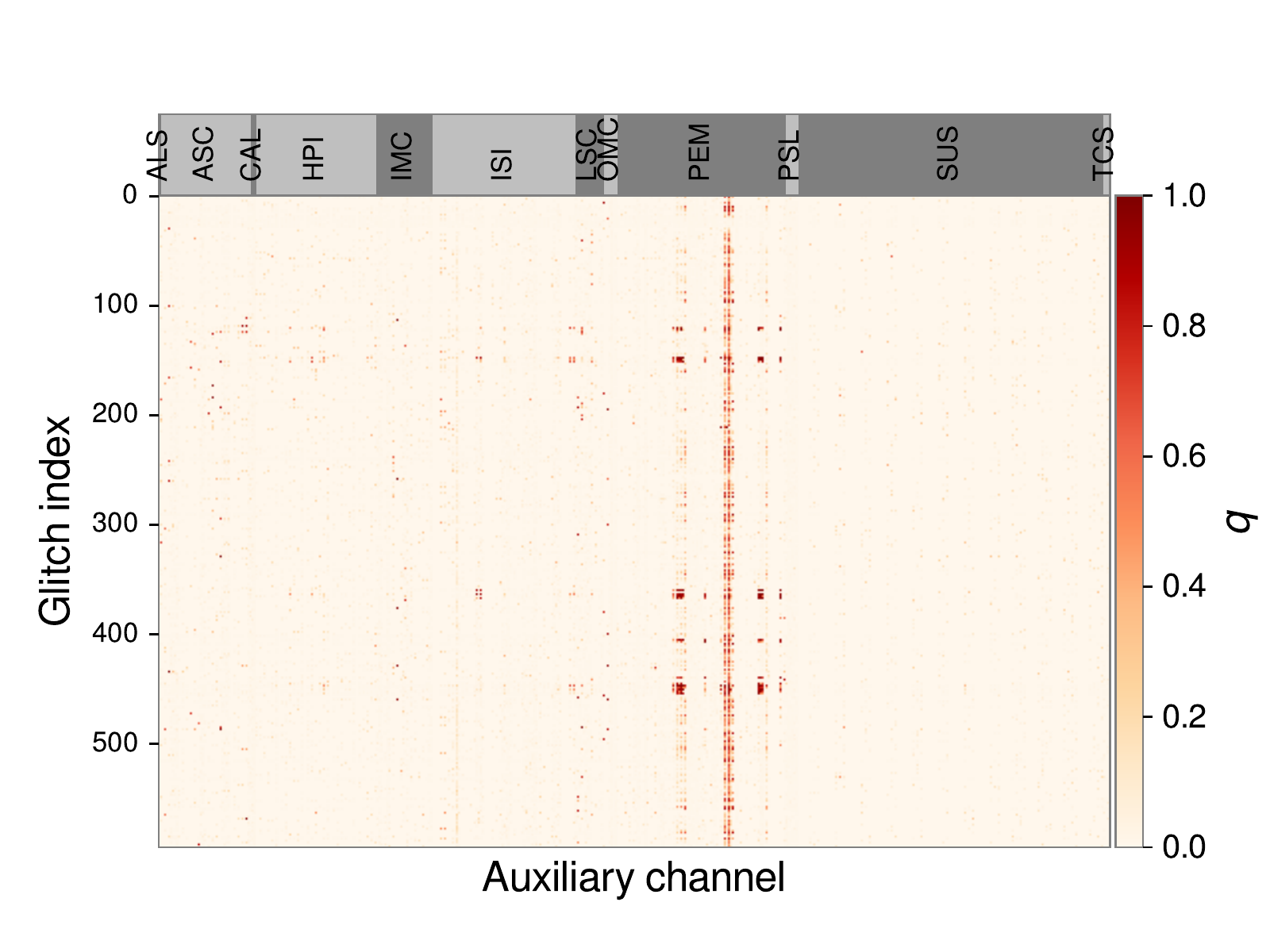}
  \end{center}
    \end{minipage}
    \begin{minipage}{0.5\hsize}
  \begin{center}
   \includegraphics[width=1\linewidth]{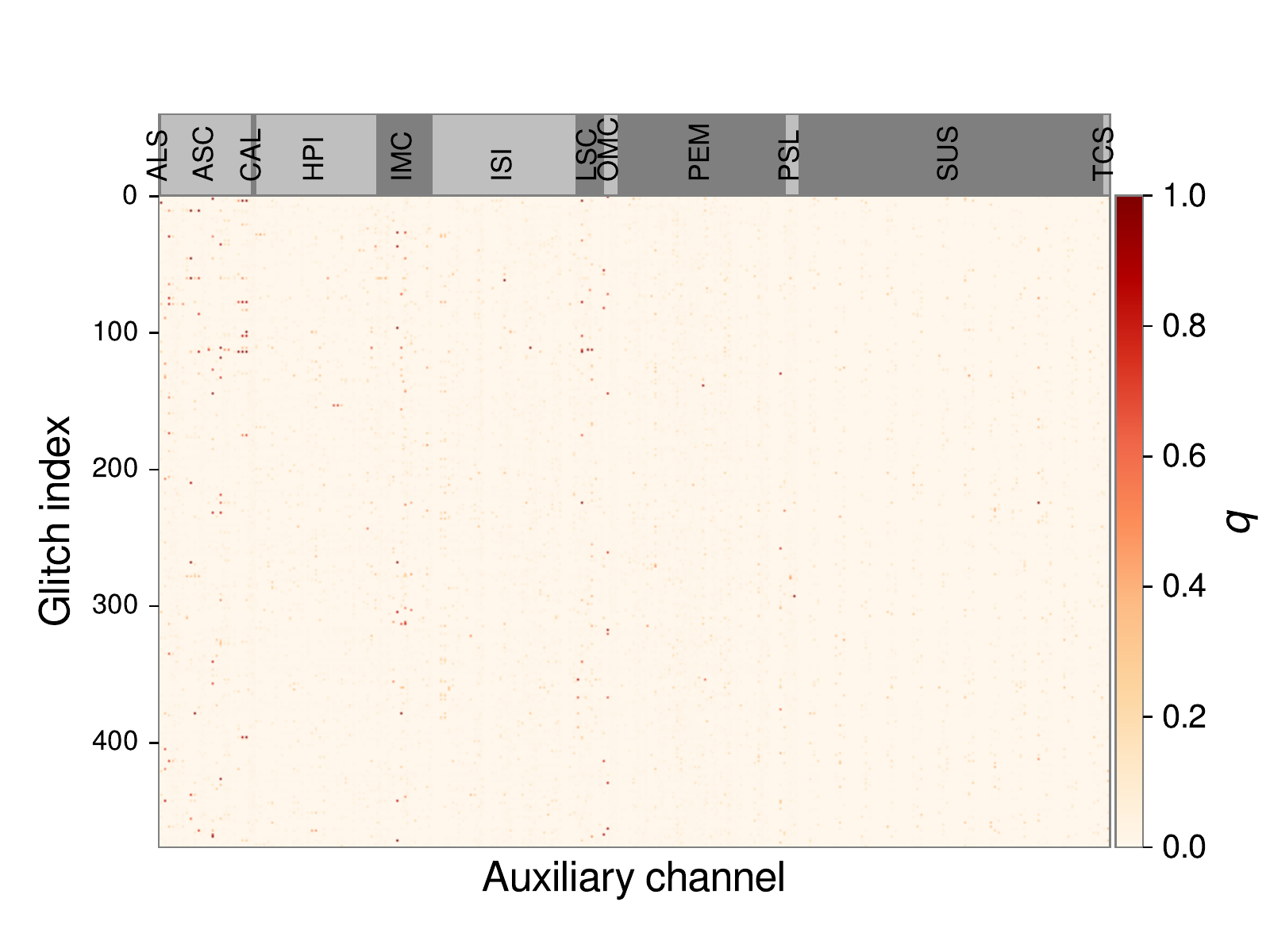}
  \end{center}
    \end{minipage}
\caption{Left: Values of the excess power measure $q$ for \secondhannel safe auxiliary channels with 8 different frequency bands in coincidence with \targetnumsecond magnetometer glitch set with \ac{SNR} $\geq 7.5$. Right: values of $q$ for the null set with a sample size of \nullnumsecond. The channels in the 8 different frequency bands are shown side by side from left to right where the chosen frequency bands are 1-128 Hz, 128-256 Hz, 256-512 Hz, 512-1024 Hz, 1024-2048 Hz, 2048-4096 Hz, 4096-8192 Hz, and the range from 1 Hz to the Nyquist frequency. The dark and light gray bars in the top denote each of the channel groups in the common sub-instrumental sensor or environmental monitor.}
\label{fig:glich_index_vs_channel}
\end{figure}

As the target set, we choose \targetnumsecond glitch samples with \ac{SNR} $\geq 7.5$ from this {\it magnetometer} set. Also, we create the null set with a sample size of \nullnumsecond by analyzing the randomly chosen timestamps when no Omicron triggers with \ac{SNR} $\geq 5.5$ are present in the detector's output. We analyze \secondhannel safe auxiliary channels with 8 different frequency bands. Figure \ref{fig:glich_index_vs_channel} shows values of $q$ for the target and null sets. One of the EX magnetometer channels called PEM-EX\_MAG\_VEA\_FLOOR\_X\_DQ in the PEM sub-system has values of $q\geq 0.6$ for 93\% of the target set and 0\% of the null set. The Alignment Sensing and Control (ASC) channels show random fluctuating values of $q$ both in the target and null sets.         

\begin{figure}[ht!]
    \centering
    \includegraphics[width=\linewidth]{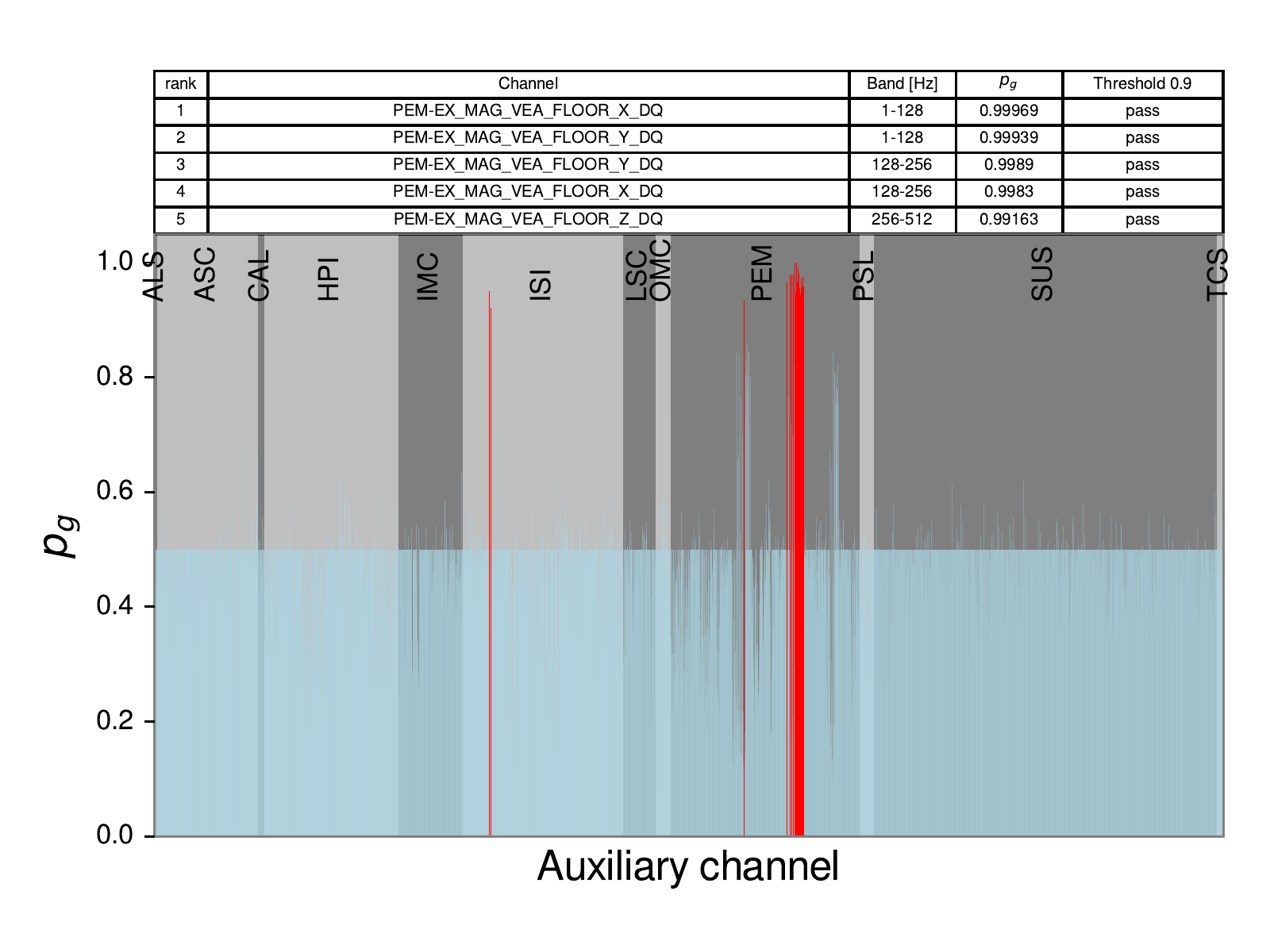}
    \caption{$p_g$ obtained from the 700 safe auxiliary channels in 8 different frequency bands for the magnetometer set. The red (cyan) bars denote channels with a given of frequency band with $p_g \geq 0.9\, (< 0.9)$. The channel in the 8 different frequency bands are shown side by side from left to right where the chosen frequency bands are 1-128 Hz, 128-256 Hz, 256-512 Hz, 512-1024 Hz, 1024-2048 Hz, 2048-4096 Hz, 4096-8192 Hz, and the range from 1 Hz to the Nyquest frequency. The dark and light gray background colors denote each of the channel groups in the common sub-instrumental sensor or environmental monitor. The table shows the top five channels in different frequency bands. \textit{Pass} (\textit{fail}) in the last column in the table show whether values of $p_g$ are above (below) 0.9.}
    \label{fig:p_g_magnetometer} 
\end{figure}

Using the target and null sets in Fig \ref{fig:glich_index_vs_channel}, we calculate the probability $p_g$ in Eq. (\ref{eq:p_belong}) for channels in each frequency band. Figure \ref{fig:p_g_magnetometer} shows values of $p_g$ for the magnetometer set. \tool successfully identifies the witness channels including the EX magnetometer channels as well as ISI-ETMX\_ST1\_BLND\_Z\_T240\_CUR\_ IN1\_DQ channel in 1-128 Hz with $p_g = 0.95$, in agreement with Ref. \cite{Cavaglia:2018xjq} . 

\section{Application to an unmodeled \ac{GW} detection pipeline} \label{result}

In \ac{GW} signal searches, the detection pipelines generate triggers with \acp{RS} (e.g., \ac{SNR}). Typically, triggers from astrophysical signals have \ac{SNR} $\gtrsim 8$. The confidence in detecting astrophysical signals is characterized by a \ac{FAR}, which is the rate of terrestrial-noise triggers with \acp{RS} equal or higher than the \ac{RS} of an astrophysical candidate event. The \ac{FAR} is typically required to be smaller than 2.0 per year \cite{Abbott:2020niy}. For increasing confidence in \ac{GW} detections, it is crucial to reduce outlying noise triggers with large values of \ac{RS}. As mentioned earlier, unmodeled detection pipelines are typically susceptible to glitches, causing a large number of noise triggers with high values of \ac{RS}. Hence, we focus on noise triggers generated with one of the unmodeled pipelines called \ac{cWB} \cite{Klimenko:2008fu,Klimenko:2015ypf} in the analysis.

We use a set of \textit{background-mode} \ac{cWB} triggers created from the data of the \ac{L1} and \ac{H1} detectors between January 1st, 2020 and February 3rd, 2020. Because the background-mode \ac{cWB} applies some time shifts much longer than the light-travel time between detectors, these triggers represent noise artifacts. In our analysis, there is \textit{no} trigger of astrophysical signals in origin. In this period, the detector's output was significantly contaminated with glitches, resulting in 40 \ac{cWB} triggers with the \ac{RS} of $\rho>9$ being generated. Around the trigger time in the \ac{L1} detector, we analyze the data from auxiliary channels with \tool.

Because more than one trigger representing a same glitch could be generated in the proximity of trigger times in a detector, we cluster the \ac{cWB} triggers by keeping the subset of triggers with the largest value of $\rho$ in the window of 0.5 seconds to avoid double-counting glitches. For the target set, we choose the \targetnumthird clustered outlying trigger with $\rho>9$. We consider the clustered trigger times as the center times for the target samples. Also, we manually choose the duration of 1 second for these samples because the durations provided by cWB are found to be too small (typically $\sim 0.01$ seconds) to represent the durations of glitches. For the null set, we create \nullnumthird samples with a duration of 1 second during the quiet period when there is no Omicron trigger with \ac{SNR} $\geq 5.5$. We use the \thirdchannel safe channels in the \ac{L1} detector and 9 different frequency bands for each of the channels. In this analysis, we have a band of 1-50 Hz in addition to a list of bands used in the previous section because the peak frequency of some \ac{cWB} triggers is around 20 Hz and the witness of those triggers are expected to have excess power only in the low-frequency (below $\sim40$ Hz) region.  

\begin{figure}[ht!]
    \centering
    \includegraphics[width=\linewidth]{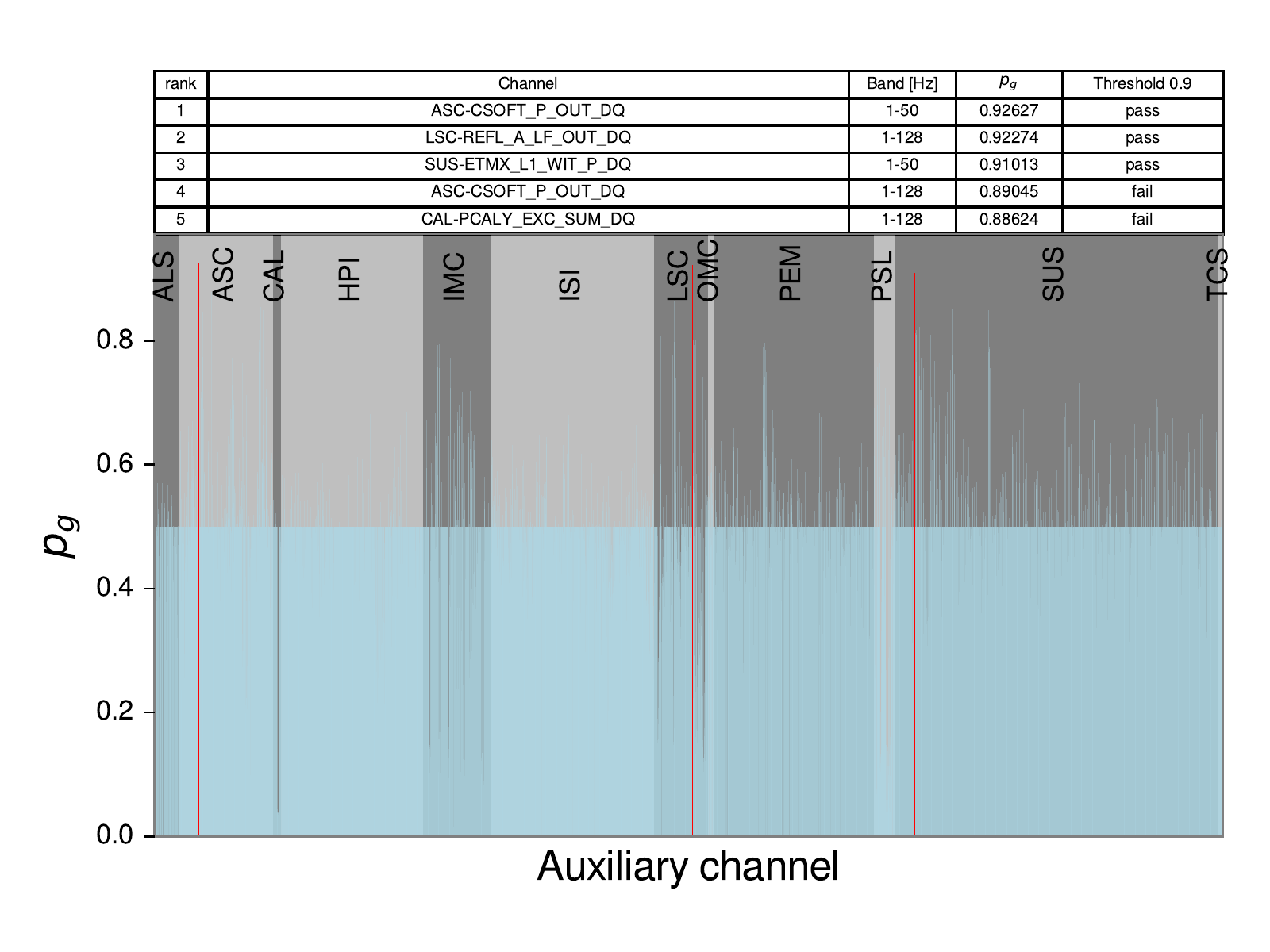}
    \caption{$p_g$ obtained from the \thirdchannel safe auxiliary channels in 8 different frequency bands for the clustered outlying cWB triggers with $\rho > 9$ from January 1st to February 3rd, 2020. The red (cyan) bars denote channels with a given of frequency band with $p_g \geq 0.9\, (< 0.9)$. The channels in the 9 different frequency bands are shown side by side from left to right where the chosen frequency bands are 1-50 Hz, 1-128 Hz, 128-256 Hz, 256-512 Hz, 512-1024 Hz, 1024-2048 Hz, 2048-4096 Hz, 4096-8192 Hz, and the range from 1 Hz to the Nyquist frequency. The dark and light gray background colors denote each of the channel groups in the common sub-instrumental sensor or environmental monitor. The table shows the top five channels in different frequency bands. \textit{Pass} (\textit{fail}) in the last column in the table show whether values of $p_g$ are above (below) 0.9.}
    \label{fig:p_g} 
\end{figure}

Figure \ref{fig:p_g} shows three witness channels with values of $p_g \geq 0.9$. The first-ranked channel (ASC-CSOFT) monitors the motions of mirrors in the arm in the \ac{GW} detector. The second-ranked channel (LSC-REFL) typically monitors the laser intensity dips. The third-ranked channel (SUS-ETMX) monitors the displacement in the suspension system in the end station in the X-arm.  

\begin{figure}[ht!]
    \begin{minipage}{0.5\hsize}
  \begin{center}
   \includegraphics[width=1\linewidth]{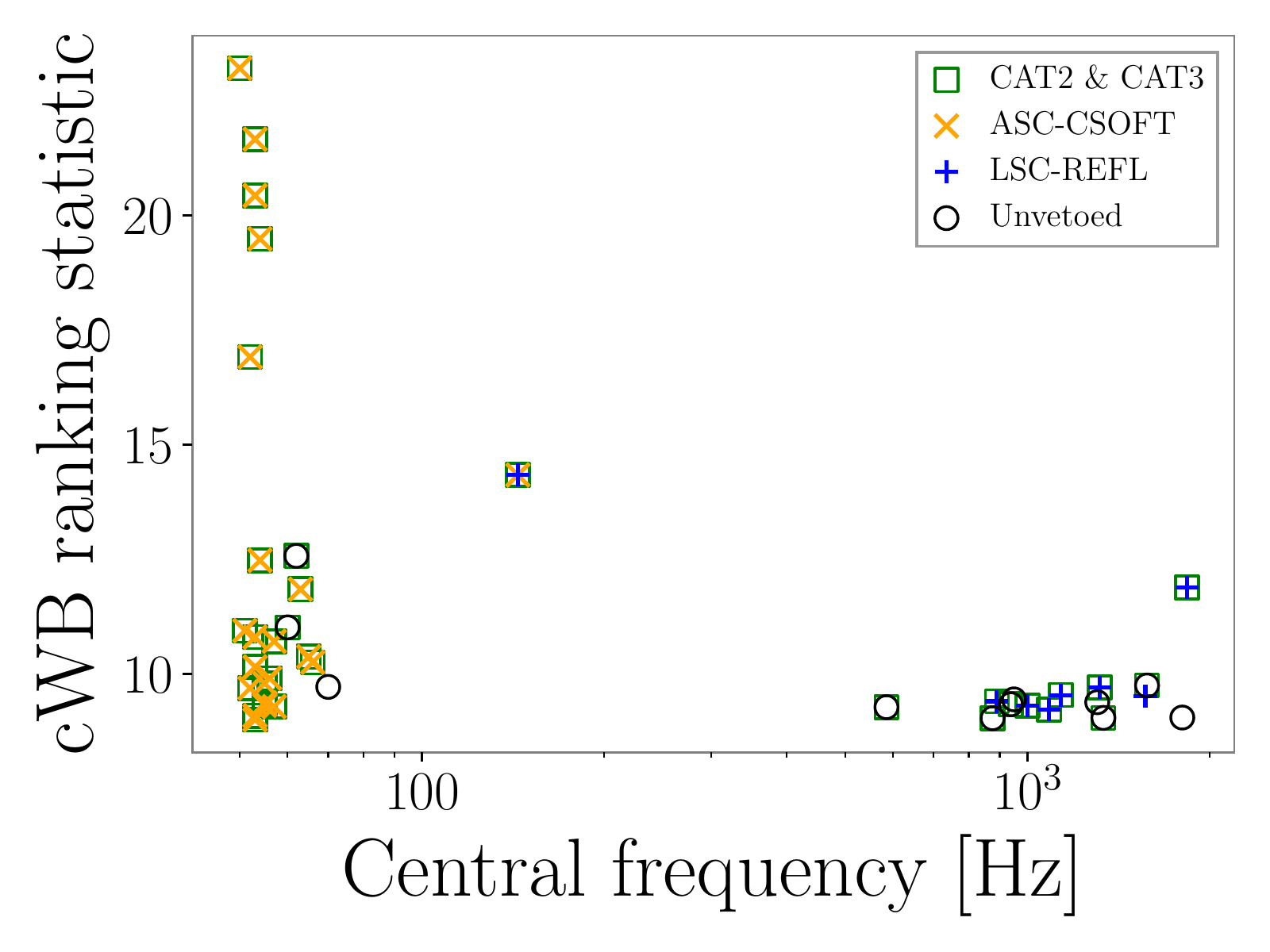}
  \end{center}
    \end{minipage}
    \begin{minipage}{0.5\hsize}
  \begin{center}
   \includegraphics[width=1\linewidth]{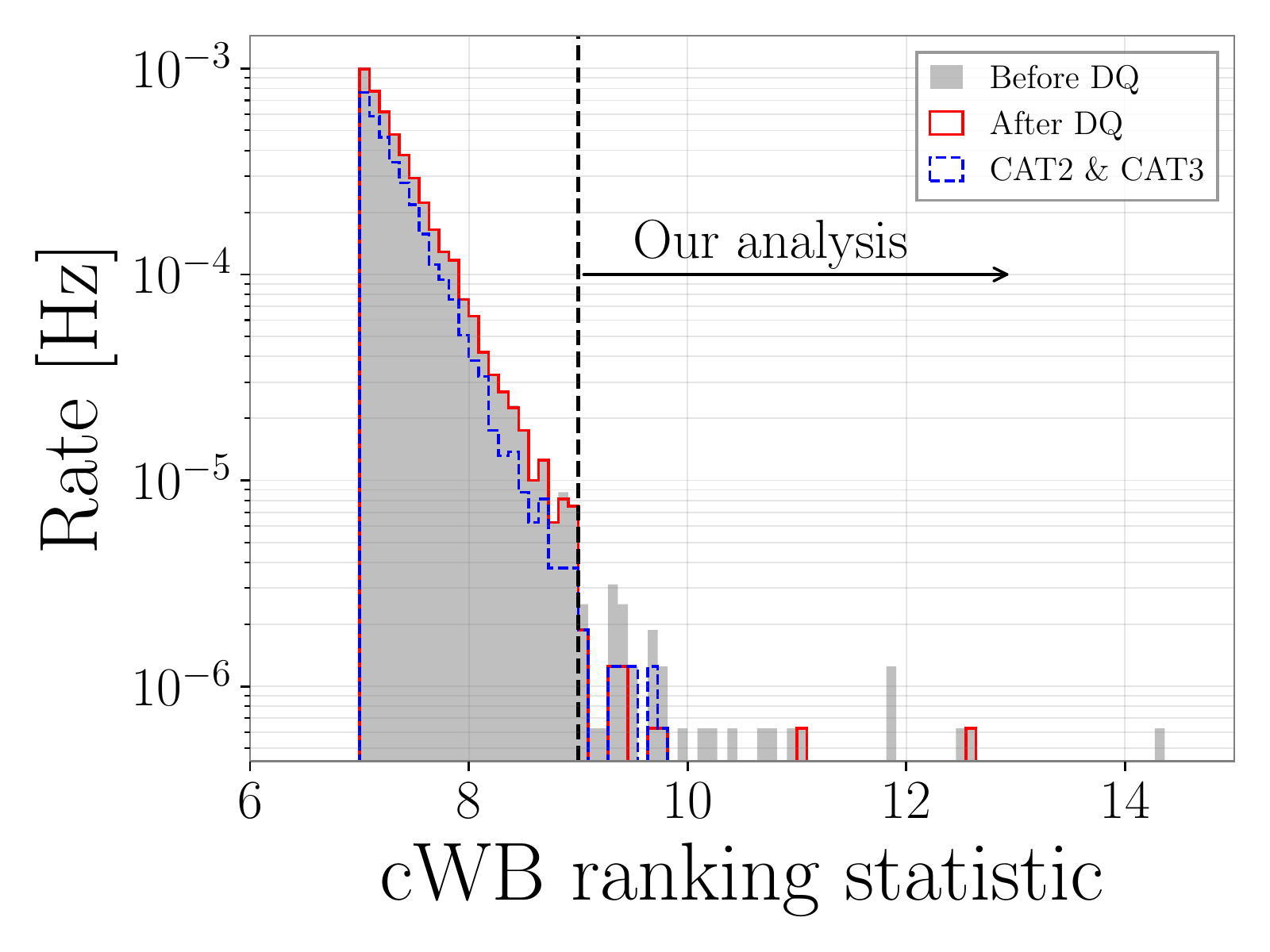}
  \end{center}
    \end{minipage}
\caption{Left: 40 outlying \textit{background} cWB triggers with the ranking statistic of $\rho > 9$. The orange-$\times$ (blue-$+$) markers denote the triggers vetoed with ASC-CSOFT (LSC-REFL) channels based on the criterion of $p_v > 0.95$. The black-circles denote the remaining triggers in our analysis. The green-squares denote the triggers vetoed with the CAT2 and CAT3 flags. Right: Rate of all the \textit{background} cWB triggers between January 1st, 2020 and February 3rd, 2020. The grey and red histograms denote the rates with cWB triggers before and after our veto, respectively. The blue-dashed histogram is the rate after the CAT and CAT3 flags applied.}
\label{fig:vetoed_trig}
\end{figure}

\begin{figure}[ht!]
    \begin{minipage}{0.5\hsize}
  \begin{center}
   \includegraphics[width=1\linewidth]{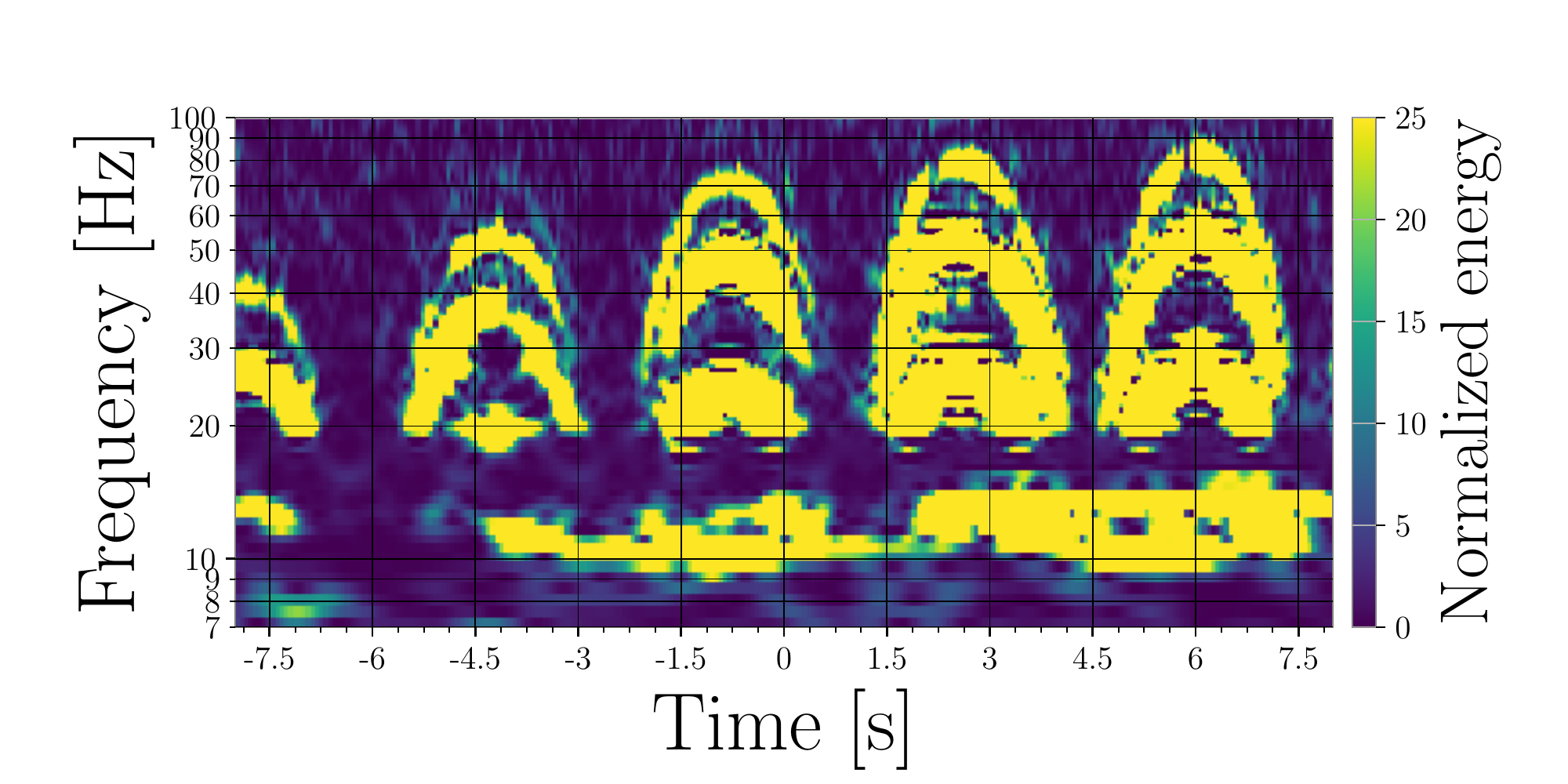}
  \end{center}
    \end{minipage}
    \begin{minipage}{0.5\hsize}
  \begin{center}
   \includegraphics[width=1\linewidth]{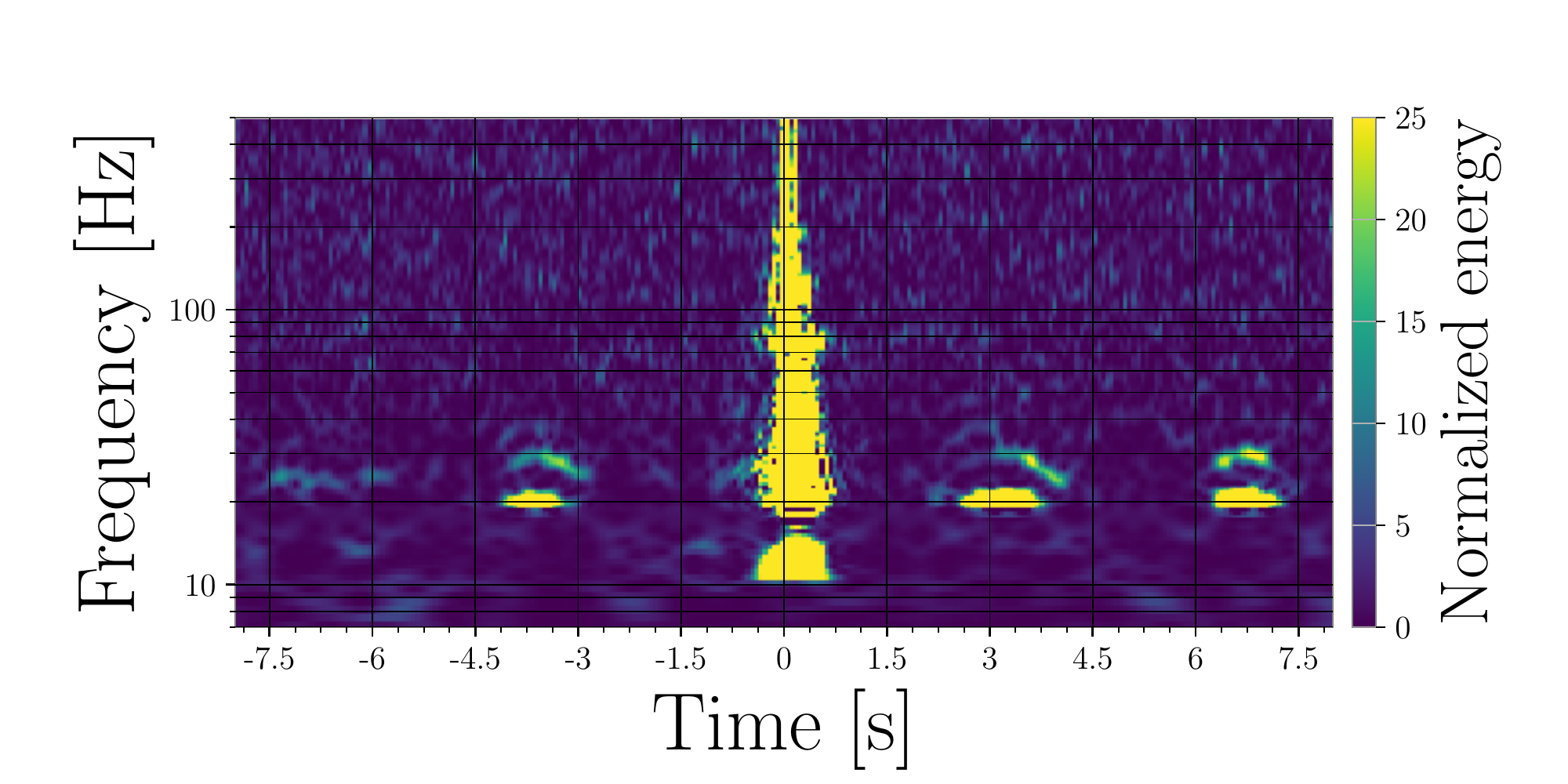}
  \end{center}
    \end{minipage}
\begin{minipage}{0.5\hsize}
  \begin{center}
   \includegraphics[width=1\linewidth]{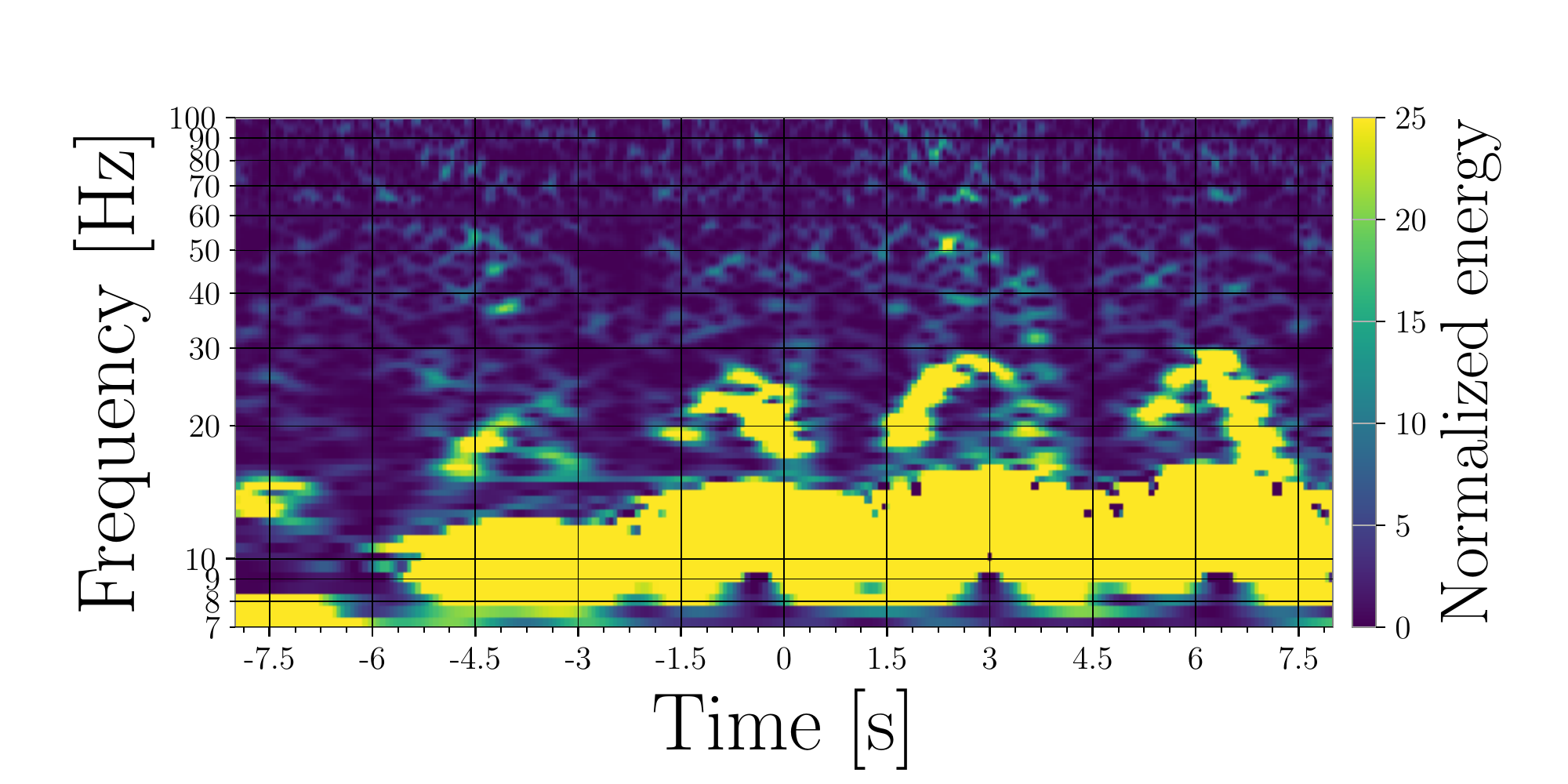}
  \end{center}
    \end{minipage}
    \begin{minipage}{0.5\hsize}
  \begin{center}
   \includegraphics[width=1\linewidth]{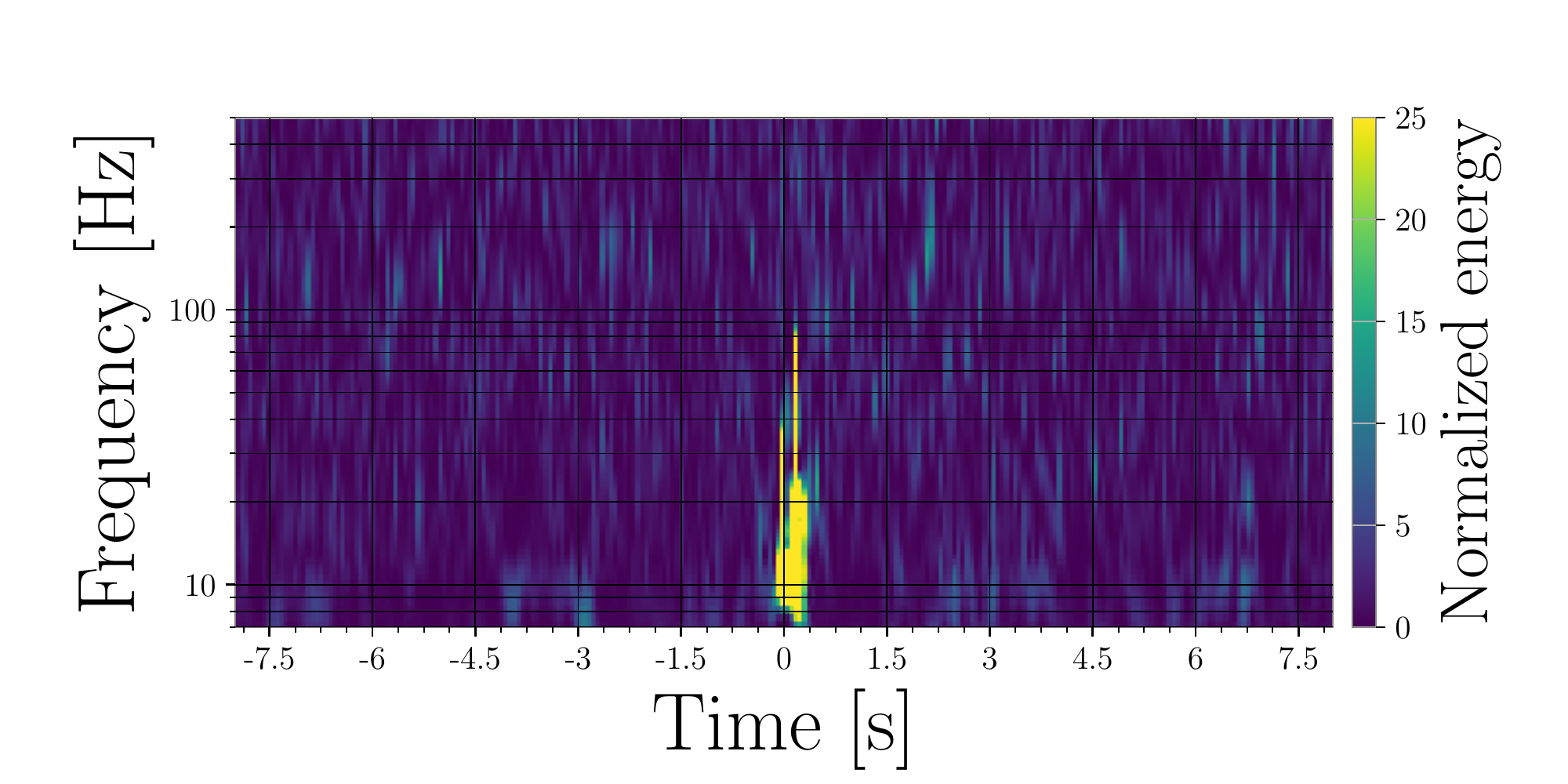}
  \end{center}
    \end{minipage}
\caption{The time-frequency representation of the glitches in the detector's output (top) and the excess power in the witness auxiliary channels (bottom) using the Q-transform \cite{ROBINET2020100620}. The left and right panels show the glitches witnessed with the ASC-CSOFT and LSC-REFL channels, respectively. The trigger times are marked as zero on the maps. The ASC-CSOFT channel in 1-50 Hz and and the LSC-REFL channel in 1-128 Hz have the values of $p_v=0.999$ and $p_v=0.954$ for each glitch, respectively.}
\label{fig:freq-time}
\end{figure}

Using values of $p_v$ obtained with the top two channels in the frequency band in Fig. \ref{fig:p_g}, we consider triggers to be vetoed using the 1-second window around the trigger. We find that the channels up to the second-ranked are sufficient because no additional triggers can be removed by adding the third-ranked channel. For vetoing triggers, we choose a conservative criterion of $p_v > 0.95$. The left panel in Fig. \ref{fig:vetoed_trig} shows the cWB outlying triggers which can be vetoed. The triggers with a central frequency less than 80 Hz are typically vetoed with the ASC-CSOFT channel because the mirror motion produces low-frequency glitches shown in the left panels in Fig. \ref{fig:freq-time}. Because the glitches produced by the laser power intensity dips typically have a large bandwidth ranging from $\sim$10 to $\sim$ 2000 Hz, vetoed triggers with the LSC-REFL have the central frequencies of either less than $\sim110$ Hz or greater than $\sim800$ Hz. Figure \ref{fig:freq-time} shows two representative glitches witnessed with either the ASC-CSOFT or LSC-REFL channels. Overall, 72.5\% of 40 outlying triggers in this search period can be vetoed with our analysis. The right panel in Fig. \ref{fig:vetoed_trig} shows that the rates of $\rho$ in the cWB triggers before and after the veto. 

As a complementary check, we compare our veto performance with that obtained by the current LIGO infrastructure. Using a set of veto periods obtained from three different flag categories indicating: 1) a critical issue with an abnormally operating detector (CAT1) ; 2) times of glitches with understood physical coupling between auxiliary channels and the detector's output (CAT2); and 3) times of glitches with unknown causes but statistical correlation (CAT3), 35 out of the cWB outliers can be vetoed. Because the periods of the CAT1 flag are already removed commonly for the analyses, we compare unvetoed triggers between our analysis and the union of the CAT2 and CAT3 flags. There are 4 commonly unvetoed triggers. We veto a single trigger that is not vetoed with the CAT2 and CAT3 flags because of the chance-coincident excess power witnessed with the LSC-LEFT channel. Seven triggers are vetoed with the CAT2 and CAT3 flags but not vetoed with our analysis. This discrepancy can be explained for two reasons. Because our veto window is 1 second around the trigger time, the quantity for the excess power outside this window is not large enough to pass our veto criterion. Otherwise, channels other than the two high-ranked witness channels selected in our analysis might witness coincident excess power. Table \ref{table:unvetoed_trig} summarizes details about these triggers. In conclusion, we have 80\% of 40 triggers are in common between our analysis and the CAT2 and CAT3 flags.

\section{Conclusion} \label{discussion_conclusion}
In this paper, we have presented a new software, \tool, designed to identify the origin of glitches and remove the effect of glitches.

Using a set of time series recorded in the instrumental and environmental monitors which do not causally follow from the detector's output, \tool queries the time series from each of the sensors around the times of glitches and then counts the fraction ($q$) of frequency bins above the stationarity upper threshold to quantify excess power. Comparing with another data set when the detector's output is quiet that is analyzed in the same way, the witness sensors are probabilistically identified based on values of $q$. To remove the effect of the glitches in the detector's output, time periods when the witness sensors monitor excess power in coincidence with the glitches, are marked as a veto. The veto criterion is given as a probability that a value of $q$ obtained with a witness channel belongs to the distribution of $q$ in the glitch set.

To demonstrate the effect on \ac{GW} searches, we have used the \textit{background} triggers given by the \ac{cWB} pipeline running on the \ac{L1} and \ac{H1} detectors from January 1st, 2020 to February 3rd, 2020. Because these triggers are generated by applying some time shifts longer than the light-travel time between detectors, there is \textit{no} trigger of astrophysical signals in origin in our analysis. During this period, the detector's output was significantly contaminated with glitches. We analyze the data of the \ac{L1} auxiliary channels using the \ac{L1} trigger time of the outlying triggers with the ranking statistic of $\rho>9$. We find at least two kinds of adversely affecting glitches to the pipeline, one of which seems to be due to the mirror-motions witnessed by the ASC-CSOFT channel and the other one seems to be typically caused by the laser intensity dips witnessed by the LSC-REFL channel with high confidence. Using these two witness channels, we consider that 72.5\% of 40 outlying triggers to be vetoed. We find that none of the cWB triggers marked as being vetoed are in coincidence with super events reported in the database server of the candidate \ac{GW} events (GraceDB) \cite{GraceDB}.

As a complementary check, we compare our results with the current LIGO infrastructure; the CAT2 and CAT3 flags. We veto a single trigger that is not vetoed with the CAT2 and CAT2 flags because of the coincident excess power witness with the LSC-LEFT channel. Our analysis does not veto 7 triggers that are vetoed with the CAT2 and CAT3 flags because the excess power is present outside of the 1-second window used in our analysis or the coincident excess power is not witnessed with our selected two channels. Overall, 80\% of the triggers are in common between our analysis and the LIGO infrastructure.

As mentioned, these 40 outlying triggers seem to have at least two distinct sources of glitches. In our analysis, we have used all triggers to calculate $p_v$. Values of $p_v$ could be higher by grouping glitch samples based on $q$ of all channels and calculate $p_v$ for each group. To group samples, machine-learning clustering algorithms such as \textsc{Gaussian Mixture Clustering} \cite{Ghosh1984OnTA, Hartigan1985AFO} or \textsc{Agglomerative Clustering} \cite{10.2307/2346439} can be applied after using some dimensionality reduction algorithms including \textsc{\ac{PCA}} \cite{doi:10.1080/14786440109462720, hotelling1933analysis, Minka00automaticchoice} in \textsc{Scikit-Learn} \cite{scikit-learn}. \tool has the in-progress implementation using \textsc{Gaussian Mixture Clustering} and \ac{PCA} incorporating statistical tests such as a one-sided binomial test and a one-sided Welch's t-test \cite{10.1093/biomet/34.1-2.28} to determine the number of clusters. Also, because the background \ac{cWB} triggers are generated by applying some time shifts between detectors, our analysis might have a bias due to one realization of time shifts. To reduce the bias, a higher number of time shifts can be chosen to generate a larger number of background \ac{cWB} triggers.

\tool has several advantages. Firstly, it can work with any \acp{ETG} running only on the detector's output without the use of them running on auxiliary channels. Secondly, because a list of glitches can be chosen by a user, it can be used to help to understand the origin of glitches which are only adversely affecting a particular \ac{GW} detection pipeline with specific parameters, e.g., high ranking statistic. The current existing veto infrastructure typically uses Omicron \ac{ETG}. Omicron is intended to capture a wide variety of glitches including those that are not similar to astrophysical \ac{GW} signals. Analyzing only the data set obtained from a particular \ac{GW} detection pipeline might make improvements in reducing veto times. Also, it takes less than 1 minute to analyze all the safe auxiliary channels and identify potentially witness channels for a given trigger. Therefore, it can be used for medium latency operations to assess if a trigger is due to astrophysical or terrestrial origin.

In addition to the purpose of vetoing glitches and the usage for the medium latency operations, \tool has another crucial advantage. As discussed in Sec. \ref{validation}, \tool can find witness channels that \ac{hVETO} might miss so that it can be used for the follow-up study about glitches identified by \ac{hVETO}. This feature is beneficial for understanding more thoroughly about noise couplings inside the instruments. For glitches with the instrumental origin, noise couplings that cause glitches could be potentially mitigated by tuning the pieces of equipment' setting or replacing them with improved ones. If mitigating of the cause is difficult to operate or the cause is of environmental origin, we envision that those glitches can be subtracted using the data recorded in witness channels based on a method similar to Ref. \cite{Ormiston:2020ele} but adapted to transient noise artifacts. The Bayesian inference approach to subtract glitches is available for a signal from the compact binary merger \cite{Chatziioannou:2021ezd, Cornish:2021wxy}. However, a subtraction method using auxiliary channels could have a significant impact on unmodeled search pipelines in the future.   

As the detector's sensitivity increases, in particular, at the low-frequency region below $\sim$ 80 Hz, unmodeled \ac{GW} detection pipelines play important roles in observing \ac{IMBBH} following the detection on May 21$^{\rm st}$, 2019 \cite{Abbott:2020tfl}. Understanding the cause of glitches and mitigating those will be more crucial.

%
%
%
%
%


\section*{Acknowledgements}

K.M. is supported by the U.S.\ National Science Foundation grant PHY-1921006 and PHY-2011334. The authors would like to thank their LIGO Scientific Collaboration and Virgo Collaboration colleagues for their help and useful comments, in particular Amber L. Stuver, Duncan MacLeod, Brennan Hughey, Marco Cavagli\`a, and Ryan Quitzow-James. The authors are grateful for computational resources provided by the LIGO Laboratory and supported by the U.S.\ National Science Foundation Grants PHY-0757058 and PHY-0823459, as well as a service of the LIGO Laboratory, the LIGO Scientific Collaboration and the Virgo Collaboration. LIGO was constructed and is operated by the California Institute of Technology and Massachusetts Institute of Technology with funding from the U.S.\ National Science Foundation under grant PHY-0757058. Virgo is funded by the French Centre National de la Recherche Scientifique (CNRS), the Italian Istituto Nazionale di Fisica Nucleare (INFN) and the Dutch Nikhef, with contributions by Polish and Hungarian institutes. The publicly available source code in \href{https://git.ligo.org/kentaro.mogushi/origli}{GitLab} is made use of python packages including \textsc{Scipy} \cite{2020SciPy-NMeth}, \textsc{GWpy}  \cite{duncan_macleod_2020_4301851}, \textsc{Pandas} \cite{reback2020pandas}, , \textsc{GWtrigfind} \cite{gwtrigfind}, \textsc{Matplotlib} \cite{Hunter:2007}, \textsc{nds2utils} \cite{nds2utils}, and \textsc{scikit-learn} \cite{scikit-learn}. This manuscript has been assigned LIGO Document Control Center number LIGO-P2100031.

\section*{References}
\bibliography{BibFile}

\providecommand{\newblock}{}
\begin{thebibliography}{10}
\expandafter\ifx\csname url\endcsname\relax
  \def\url#1{{\tt #1}}\fi
\expandafter\ifx\csname urlprefix\endcsname\relax\def\urlprefix{URL }\fi
\providecommand{\eprint}[2][]{\url{#2}}

\bibitem{Abbott:2016GW150914}
Abbott B~P {\em et~al.\/} (Virgo, LIGO Scientific) 2016 {\em Phys. Rev.
  Lett.\/} {\bf 116} 061102 (\textit{Preprint} \eprint{1602.03837})

\bibitem{TheLIGOScientific:2014jea}
Aasi J {\em et~al.\/} (LIGO Scientific) 2015 {\em Class. Quant. Grav.\/} {\bf
  32} 074001 (\textit{Preprint} \eprint{1411.4547})

\bibitem{TheVirgo:2014hva}
Acernese F {\em et~al.\/} (VIRGO) 2015 {\em Class. Quant. Grav.\/} {\bf 32}
  024001 (\textit{Preprint} \eprint{1408.3978})

\bibitem{LIGOScientific:2018mvr}
Abbott B {\em et~al.\/} (LIGO Scientific, Virgo) 2019 {\em Phys. Rev. X\/} {\bf
  9} 031040 (\textit{Preprint} \eprint{1811.12907})

\bibitem{Abbott:2020niy}
Abbott R {\em et~al.\/} (LIGO Scientific, Virgo) 2020  (\textit{Preprint}
  \eprint{2010.14527})

\bibitem{Isogai:2010zz(UPV)}
Isogai T (LIGO Scientific, Virgo) 2010 {\em J. Phys. Conf. Ser.\/} {\bf 243}
  012005

\bibitem{Smith:2011an(hVETO)}
Smith J~R, Abbott T, Hirose E, Leroy N, Macleod D, McIver J, Saulson P and
  Shawhan P 2011 {\em Class. Quant. Grav.\/} {\bf 28} 235005 (\textit{Preprint}
  \eprint{1107.2948})

\bibitem{Essick:2020cyv}
Essick R, Mo G and Katsavounidis E 2020  (\textit{Preprint}
  \eprint{2011.13787})

\bibitem{Biswas:2013wfa}
Biswas R {\em et~al.\/} 2013 {\em Phys. Rev. D\/} {\bf 88} 062003
  (\textit{Preprint} \eprint{1303.6984})

\bibitem{Klimenko:2008fu}
Klimenko S, Yakushin I, Mercer A and Mitselmakher G 2008 {\em Class. Quant.
  Grav.\/} {\bf 25} 114029 (\textit{Preprint} \eprint{0802.3232})

\bibitem{Klimenko:2015ypf}
Klimenko S {\em et~al.\/} 2016 {\em Phys. Rev. D\/} {\bf 93} 042004
  (\textit{Preprint} \eprint{1511.05999})

\bibitem{Sutton:2009gi}
Sutton P~J {\em et~al.\/} 2010 {\em New J. Phys.\/} {\bf 12} 053034
  (\textit{Preprint} \eprint{0908.3665})

\bibitem{Nitz:2017svb}
Nitz A~H, Dent T, Dal~Canton T, Fairhurst S and Brown D~A 2017 {\em Astrophys.
  J.\/} {\bf 849} 118 (\textit{Preprint} \eprint{1705.01513})

\bibitem{Sachdev:2019vvd}
Sachdev S {\em et~al.\/} 2019  (\textit{Preprint} \eprint{1901.08580})

\bibitem{ROBINET2020100620}
Robinet F, Arnaud N, Leroy N, Lundgren A, Macleod D and McIver J 2020 {\em
  SoftwareX\/}  100620 ISSN 2352-7110
  \urlprefix\url{http://www.sciencedirect.com/science/article/pii/S2352711020303332}

\bibitem{Nitz:2018rgo}
Nitz A~H, Dal~Canton T, Davis D and Reyes S 2018 {\em Phys. Rev. D\/} {\bf 98}
  024050 (\textit{Preprint} \eprint{1805.11174})

\bibitem{Zevin:2016qwy}
Zevin M {\em et~al.\/} 2017 {\em Class. Quant. Grav.\/} {\bf 34} 064003
  (\textit{Preprint} \eprint{1611.04596})

\bibitem{Cutler:1994ys}
Cutler C and Flanagan E~E 1994 {\em Phys. Rev. D\/} {\bf 49} 2658--2697
  (\textit{Preprint} \eprint{gr-qc/9402014})

\bibitem{2020SciPy-NMeth}
Virtanen P, Gommers R, Oliphant T~E, Haberland M, Reddy T, Cournapeau D,
  Burovski E, Peterson P, Weckesser W, Bright J, {van der Walt} S~J, Brett M,
  Wilson J, Millman K~J, Mayorov N, Nelson A~R~J, Jones E, Kern R, Larson E,
  Carey C~J, Polat {\.I}, Feng Y, Moore E~W, {VanderPlas} J, Laxalde D,
  Perktold J, Cimrman R, Henriksen I, Quintero E~A, Harris C~R, Archibald A~M,
  Ribeiro A~H, Pedregosa F, {van Mulbregt} P and {SciPy 10 Contributors} 2020
  {\em Nature Methods\/} {\bf 17} 261--272

\bibitem{Cavaglia:2018xjq}
Cavaglia M, Staats K and Gill T 2019 {\em Commun. Comput. Phys.\/} {\bf 25}
  963--987 (\textit{Preprint} \eprint{1812.05225})

\bibitem{DBLP:journals/corr/abs-1708-03157}
Staats K, Pantridge E~R, Cavaglia M, Milovanov I and Aniyan A 2017 {\em CoRR\/}
  {\bf abs/1708.03157} (\textit{Preprint} \eprint{1708.03157})
  \urlprefix\url{http://arxiv.org/abs/1708.03157}

\bibitem{Breiman:2001hzm}
Breiman L 2001 {\em Machine Learning\/} {\bf 45} 5--32

\bibitem{GraceDB}
Gracedb \url{https://gracedb.ligo.org/documentation/index.html} accessed:
  2021-01-26

\bibitem{Ghosh1984OnTA}
Ghosh J and Sen P 1984 On the asymptotic performance of the log likelihood
  ratio statistic for the mixture model and related results

\bibitem{Hartigan1985AFO}
Hartigan J~A 1985 A failure of likelihood asymptotics for normal mixtures

\bibitem{10.2307/2346439}
Gower J~C and Ross G~J~S 1969 {\em Journal of the Royal Statistical Society.
  Series C (Applied Statistics)\/} {\bf 18} 54--64 ISSN 00359254, 14679876
  \urlprefix\url{http://www.jstor.org/stable/2346439}

\bibitem{doi:10.1080/14786440109462720}
FRS K~P 1901 {\em The London, Edinburgh, and Dublin Philosophical Magazine and
  Journal of Science\/} {\bf 2} 559--572 (\textit{Preprint}
  \eprint{https://doi.org/10.1080/14786440109462720})
  \urlprefix\url{https://doi.org/10.1080/14786440109462720}

\bibitem{hotelling1933analysis}
Hotelling H 1933 {\em Journal of Educational Psychology\/}  417--441

\bibitem{Minka00automaticchoice}
Minka T~P 2000 Automatic choice of dimensionality for pca Tech. rep.

\bibitem{scikit-learn}
Pedregosa F, Varoquaux G, Gramfort A, Michel V, Thirion B, Grisel O, Blondel M,
  Prettenhofer P, Weiss R, Dubourg V, Vanderplas J, Passos A, Cournapeau D,
  Brucher M, Perrot M and Duchesnay E 2011 {\em Journal of Machine Learning
  Research\/} {\bf 12} 2825--2830

\bibitem{10.1093/biomet/34.1-2.28}
WELCH B~L 1947 {\em Biometrika\/} {\bf 34} 28--35 ISSN 0006-3444
  (\textit{Preprint}
  \eprint{https://academic.oup.com/biomet/article-pdf/34/1-2/28/553093/34-1-2-28.pdf})
  \urlprefix\url{https://doi.org/10.1093/biomet/34.1-2.28}

\bibitem{Ormiston:2020ele}
Ormiston R, Nguyen T, Coughlin M, Adhikari R~X and Katsavounidis E 2020 {\em
  Phys. Rev. Res.\/} {\bf 2} 033066 (\textit{Preprint} \eprint{2005.06534})

\bibitem{Chatziioannou:2021ezd}
Chatziioannou K, Cornish N, Wijngaarden M and Littenberg T~B 2021
  (\textit{Preprint} \eprint{2101.01200})

\bibitem{Cornish:2021wxy}
Cornish N~J 2021  (\textit{Preprint} \eprint{2101.01188})

\bibitem{Abbott:2020tfl}
Abbott R {\em et~al.\/} (LIGO Scientific, Virgo) 2020 {\em Phys. Rev. Lett.\/}
  {\bf 125} 101102 (\textit{Preprint} \eprint{2009.01075})

\bibitem{duncan_macleod_2020_4301851}
Macleod D, Urban A~L, Coughlin S, Massinger T, Pitkin M, rngeorge, paulaltin,
  Areeda J, Singer L, Quintero E, Leinweber K and Badger T~G 2020 gwpy/gwpy:
  2.0.2 \urlprefix\url{https://doi.org/10.5281/zenodo.4301851}

\bibitem{reback2020pandas}
pandas~development team T 2020 pandas-dev/pandas: Pandas
  \urlprefix\url{https://doi.org/10.5281/zenodo.3509134}

\bibitem{gwtrigfind}
gwtrigfind \url{https://pypi.org/project/gwtrigfind/0.8.0/} accessed:
  2021-01-26

\bibitem{Hunter:2007}
Hunter J~D 2007 {\em Computing in Science \& Engineering\/} {\bf 9} 90--95

\bibitem{nds2utils}
nds2utils \url{https://pypi.org/project/nds2utils/} accessed: 2021-01-26

\end{thebibliography}

\begin{table}[!ht]
\renewcommand*{\arraystretch}{1}
\setlength{\tabcolsep}{2.5pt}
\begin{center}
\begin{tabular}{c c c c} \toprule

 \makecell{GPS time \\ in L1} & \makecell{ASC-CSOFT \\ in 1-50 Hz}  & \makecell{LSC-REFL \\ in 1-128 Hz} & Comments \\
\cmidrule(rl){1-1} \cmidrule(rl){2-2} \cmidrule(rl){3-3} \cmidrule(rl){4-4}

1262326892.30 & 0.874 & 0.905 & Excess power in $(-2.0, -0.7)$ seconds in LSC-RELF  \\
1262403661.68 & 0.824 & 0.936 & Excess power in 1-50 Hz in LSC-REFL with $p_v=0.953$ \\
1262655787.22 & 0.785 & 0.942 & Glitches in ($0.4, 1.0$) seconds at $\sim 25$ Hz\\
1262674230.37 & 0.874 & 0.930 & A glitch at -1.4 seconds at $\sim 12$ Hz\\
1262676098.05 & 0.785 & 0.938 & Glitches below 9 Hz\\
1262758399.380 & 0.785 & 0.950 & Quiet within (-8, 8) seconds \\
\rowcolor{Gray}
1262842889.30 & 0.824 & 0.899 & -\\
\rowcolor{Gray}
1263363175.48 & 0.874 & 0.905  & - \\
\rowcolor{Gray}
1263715708.16 & 0.941 & 0.903 & -\\
\rowcolor{Gray}
1264327099.43 & 0.785 & 0.905 & -\\
1264703139.12 & 0.785 & 0.905 & A glitch in $(-2.5, 0)$ seconds  \\
\cmidrule(rl){1-4} 
\rowcolor{Gray}
1262664659.84 & 0.824 & 0.960 & Excess power in the LSC-REFL channel \\ 

\bottomrule
\end{tabular}
\caption{List of 11 unvetoed outlying \textit{background} \ac{cWB} triggers with $\rho > 9$ in our analysis using a threshold of $p_v > 0.95$ for the first and second ranked witness channels with comments for triggers in discrepancy between our analysis and the CAT2 and CAT3 flags. The trigger in the last row is vetoed because of the chance coincident excess power witnessed by the LSC-REFL channel. The values in the columns of ASC-CSOFT and LSC-REFL denote values of $p_v$. The shaded rows denote the unvetoed triggers based on the CAT2 and CAT3 data quality flags of LIGO.}
\label{table:unvetoed_trig}
\end{center}
\end{table}

\end{document}